\documentclass[aps,prb,superscriptaddress,amsfonts,amsmath,amssymb,floatfix,reprint,longbibliography]{revtex4-2}

\usepackage{url}
\usepackage{bm}
\usepackage{graphicx}
\usepackage{amsmath}
\usepackage{subfigure}
\usepackage{amstext}
\usepackage{amssymb}
\usepackage{amsfonts}
\usepackage{amsbsy}
\usepackage{verbatim}
\usepackage{physics}
\usepackage{braket}
\usepackage{color}
\usepackage[colorlinks=true, urlcolor=blue, linkcolor=blue, citecolor=blue, pdftex]{hyperref}
\usepackage{multirow}
\usepackage{floatrow}
\usepackage{float}
\usepackage{tikz}
\usepackage{gensymb}
\usepackage{textcomp}
\usepackage{enumitem}
\usepackage[version=3]{mhchem}
\usepackage{afterpage}
\usepackage{pbox}
\usepackage{makecell}
\usepackage{dsfont}
\usepackage{physics}
\usepackage{siunitx}
\usepackage{fancyhdr}
\usepackage{stackengine,wasysym,scalerel}
\usepackage{latexsym}
\usepackage{cancel}
\usepackage{array, multirow, bigdelim, makecell, booktabs}
\usepackage[misc]{ifsym}
\usepackage[normalem]{ulem}
\usepackage{times}
\usepackage{comment}

\definecolor{darkblue}{HTML}{004D6B}
\definecolor{darkred}{HTML}{8c1515}
\definecolor{darkgreen}{HTML}{006400}
\usepackage{hyperref}
\hypersetup{
	colorlinks=true,
	urlcolor=darkred,
	citecolor=darkblue,
	linkcolor=darkred,
	breaklinks
}

\begin{document}

\title{Simulating chiral spin liquids with fermionic Projected Entangled Paired States}

\author{Sasank Budaraju}\email{budaraju@irsamc.ups-tlse.fr}
\affiliation{Laboratoire de Physique Th\'eorique, Universit\'e de Toulouse, CNRS, UPS, France}
\affiliation{Department of Physics and Quantum Centre of Excellence for Diamond and Emergent Materials (QuCenDiEM), Indian Institute of Technology Madras, Chennai 600036, India}

\author{Didier Poilblanc}\email{didier.poilblanc@irsamc.ups-tlse.fr}
\affiliation{Laboratoire de Physique Th\'eorique, Universit\'e de Toulouse, CNRS, UPS, France}

\author{Sen Niu}\email{sen.niu@irsamc.ups-tlse.fr}
\affiliation{Laboratoire de Physique Th\'eorique, Universit\'e de Toulouse, CNRS, UPS, France}
\affiliation{Department of Physics and Astronomy, California State University Northridge, California 91330, USA}

\date{\today}

\begin{abstract}

Chiral Spin Liquids (CSL) based on spin-1/2 fermionic Projected Entangled Pair States (fPEPS) are considered on the square lattice.
First, fPEPS approximants of Gutzwiller-projected Chern insulators (GPCI) are investigated by Variational Monte Carlo (VMC) techniques on finite size tori. 
We show that such fPEPS of finite bond dimension can correctly capture the topological properties of the chiral spin liquid, as the exact GPCI, with the correct topological ground state degeneracy on the torus. Further, more general fPEPS are considered and optimized (on the infinite plane) to describe the CSL phase of a chiral frustrated Heisenberg antiferromagnet. The chiral modes are computed on the edge of a semi-infinite cylinder (of finite circumference) and shown to follow the predictions from Conformal Field Theory. In contrast to their bosonic analogs the (optimized) fPEPS do not suffer from the replication of the chiral edge mode in the odd topological sector. 

\end{abstract}

\maketitle

\section{Introduction}

Commonly, a quantum spin liquid refers to a quantum state that does not break spontaneously the system symmetries, neither the space group symmetry nor the global continuous (e.g. SU(2)) symmetry, if any of those is present in the Hamiltonian. In particular, spin liquids show absence of magnetic ordering down to zero temperature.
Chiral spin liquids (CSLs) are exotic states of matter characterized, in addition to their spin liquid character, by the breaking of time-reversal ($T$) and parity ($P$) symmetries \cite{zee1989}. They also exhibit long-range topological order~\cite{wen1991gapless}. They have been encountered in several quantum spin models with $\rm SU(2)$~\cite{bauer2014chiral,wietek_csl,hickey17,csl_dm,psg_csl,vmc_j1j2jx,wietek15,hong_su2kitaev} or higher $\rm SU(N)$~\cite{chen2021abelian,hermele_sun,nataf_csl,jychen_prl,zhangsu4,yao_prr,gang_16csl,hermele_11sun} symmetry in the presence of a chiral term breaking explicitly $T$ and $P$. In some cases, $P$ and $T$ can be broken spontaneously, as demonstrated in both Hubbard/Heisenberg like models \cite{yche_csl,hu_vmc15, yche2,gong15,gong2014emergent,moat_csl,Yao2018,hubbard_csl} and Kitaev models \cite{KITAEV20062} on lattices with odd plaquettes  \cite{yaokivelson2007,prl_amorphous_csl,non_archimdedean,csl_quasicrystal, Cassella2023, exact_csl_disorder,chua_csl11,fu19}. Parent hamiltonians have also been devised for CSLs \cite{parent_ham_1,parent_ham_3, parent_ham_2}. 

Experimental efforts to identify compounds with CSL ground states are ongoing, with prominent attention devoted to spin-orbit coupled materials with Kitaev-like interaction under external magnetic fields \cite{TREBST20221,Kasahara2018,rev_knolle}.
Additionally, in Moir\'e heterostructures, the layer degree of freedom regarded as a pseudo-spin facilitates realization of pseudo-chiral spin liquids \cite{kuhlenkamp2022tunable}. Chiral spin liquids have also been proposed in cold atom platforms using Floquet driving \cite{PRXQuantum_csl_floquet} or Rydberg arrays \cite{yang2022quantum,kalinowski2023non,ohler23,weber22} where the chirality of edge modes can be detected experimentally. CSLs relevant for Rydberg atom models on the honeycomb lattice have recently been classified via a projective symmetry group approach \cite{poetri_csl23}.

As CSLs typically emerge in strongly correlated systems, numerical tools for determining phase diagrams of generic Hamiltonians and characterizing CSLs are desired. Tensor networks like Projected Entangled Pair States (PEPS) \cite{verstraete2004renormalization} are well suited to the investigation of spin liquids. Topological orders can be encoded naturally by imposing virtual gauge symmetries \cite{toriccode2006,schuch2010}.  In addition, chiral forms of PEPS can describe CSL~\cite{poilblanc2015chiral,poilblanc20162} and be used as an efficient variational scheme to attack frustrated quantum spin models hosting CSL phases~\cite{poilblanc2017,niu2022chiral}. On the other hand, infinite PEPS (iPEPS) is an ideal tool as it defines states in the thermodynamic limit directly, avoiding finite size extrapolations. 

Despite the above mentioned successes and strengths of the PEPS framework, it has remained challenging to figure out completely whether conventional bosonic PEPS can truly describe CSL. In particular, it is still unclear whether topological obstruction~\cite{wahl2013projected,dubail2015tensor} affects, in addition to small artefacts in the long distance real-space correlations (presence of a gossamer tail~\cite{poilblanc2017,hasik2020peps}), global topological properties
like (a) the topological GS degeneracy or (b) the correct conformal field theory (CFT) counting in the entanglement spectrum (ES). For the non-chiral case, PEPS is believed to be a conceptually good ansatz, topological order being encoded by gauge symmetry~\cite{schuch2010,schuch2013}.
In contrast, whether chiral PEPS give the correct topological degeneracy of the CSL is still unsettled in general due to expensive computation cost \cite{poilblanc2015chiral}, except in very rare cases where bond dimension is very small~\cite{yang2015chiral}. For example, in the case of the $\rm SU(2)_1$ CSL, although one can insert string (Wilson loop) operators in x and/or y directions, the resulting states are not linearly independent and the degeneracy should be only $2$, that has not been definitely proven in chiral (bosonic) PEPS (although results are not inconsistent with that claim)~\cite{poilblanc2015chiral}. 
In addition, simple chiral PEPS revealed a doubling of the chiral edge branch in the odd topological sector~\cite{poilblanc2015chiral,poilblanc20162} which seems to persist in the case of fully optimized wave functions for Abelian $\rm SU(N)_1$ and non-Abelian CSLs ~\cite{poilblanc2017,hasik2020peps,chen2018non,chen2020,chen2021abelian,niu2022chiral}.

In this paper, based on the recently proposed projected fermionic PEPS (fPEPS) ansatz, we show, using VMC techniques, that PEPS can represent CSLs with correct topological degeneracy.
Using this fPEPS ansatz as initial state, we further perform variational optimization to attack a frustrated $J_1-J_2-J_{\chi}$ square lattice model~\cite{Nielsen2013} in the regime of chiral spin liquid~\cite{poilblanc2017,Yang2024,Zhang2024}. The fPEPS approach has competitive energy compared to the conventional bosonic PEPS, and crucially shows correct ES degeneracy.

The rest of this paper is structured as follows: In section \ref{method}, we discuss the numerical techniques employed, including the construction of the parton ansatz and the VMC and PEPS methods. Then, in section \ref{sec3vmc} we show results of the VMC analysis of fPEPS states on finite clusters, and in section \ref{sec4ipeps} we discuss the variational optimization of the fPEPS states on the infinite plane to study the  $J_1-J_2-J_{\chi}$ model.
\section{Numerical techniques}
\label{method}

\subsection{Parton ansatze}
\label{ansatz}

In order to construct a simple CSL, we first consider a Chern insulator state with $C=1$, obtained by diagonalizing the following (free electron) Hofstadter hamiltonian on the square lattice 
\begin{equation}
    H = \sum_{\langle i j \rangle, \sigma} t_1 \chi_{ij} c^\dagger_{i \sigma} c_{j \sigma}  + \sum_{\langle \langle i k \rangle \rangle, \sigma} t_2 c^\dagger_{i \sigma} c_{k \sigma} e^{i\theta_{ik}} + \text{h.c.}
    \label{eq:Hofstadter}
\end{equation}
where $\langle i j\rangle$ ($\langle  \langle i k \rangle \rangle$) denotes nearest (next-nearest) neighbor bonds and $\sigma$ is the spin index, $\sigma = \{\uparrow, \downarrow\}$. The hoppings are identical for both spin species. We fix $t_2 = 0.5 t_1$, $\chi_{ij} = \pm 1$ to ensure a $\pi$ flux through every square plaquette and choose the complex phases $\theta_{ik}$ to obtain a $\pi/2$ flux in all triangles. In this paper we choose the gauge used in Ref. \cite{niu2023chiral} such that the unit-cell can be chosen as two nearest neighbour sites along the $x$ direction. 

The exact many-body spin wavefunction $|\varphi\rangle$ is a Gutzwiller projected Slater-determinant
\begin{equation}
    |\varphi\rangle =P_{G}\prod_{\alpha } c^{\dagger}_{\alpha,\uparrow}c^{\dagger}_{\alpha,\downarrow}|0\rangle,
    \label{eq:corr_state}
\end{equation}
where the Gutwiller-projector $\prod_i (n_{i,\uparrow} - n_{i,\downarrow})^2$ projects onto the subspace of exactly one electron per site, and  $c^{\dagger}_{\alpha,\sigma}$ correspond to single-particle states obtained from the mean-field Hamiltonian Eq. \eqref{eq:Hofstadter}. 
For Gaussian fPEPS, the set of $c^{\dagger}_{\alpha,\sigma}$ orbitals are only represented approximately due to truncation of finite bond dimension. Details on Gaussian fPEPS will be provided in subsection B.  

It is known that the resultant exact parton state is a $\rm SU(2)_1$ CSL which is equivalent to the $\nu = 1/2$ bosonic Laughlin state. On a torus, such CSL has two-fold topological degeneracy where the degenerate states can be constructed by imposing different boundary conditions on the parton wavefunctions \cite{yizhang2012}. On a cylinder, the CSL hosts chiral gapless edge states predicted by $\rm SU(2)_1$ Wess-Zumino-Witten (WZW) CFT. In the rest of our work, we dub this parton state in \eqref{eq:corr_state} the "exact CSL", in contrast to its fPEPS approximation to be discussed below.

\subsection{Construction of Gaussian fPEPS state}
To construct the Gaussian fPEPS state which is an approximation of the exact ground state of Eq. \eqref{eq:Hofstadter}, We adopt the method introduced in Refs. \cite{mortier2022tensor,li2023u}. The translation invariant many-body ansatz is parametrized by a single Gaussian tensor with four virtual indices and two physical indices corresponding to the unitcell in the Hofstadter model. In the Gaussian tensor, the virtual space dimension is defined by the number of virtual modes $M$. Each virtual fermion mode can be occupied or unoccupied, thus the bond dimension becomes $D=2^M$ for a spinless state and $D=4^M$ for spinful $\rm SU(2)$ state. To obtain the best approximation of the Gaussian fPEPS tensor, we use gradient optimization and choose the (free electron) energy of Eq. \eqref{eq:Hofstadter} at half-filling as cost function. 

As the unprojected fPEPS state is Gaussian, it can be also written as a Slater-determinant (product state) on any finite torus and all the physical properties can be extracted exactly. In Ref. \cite{niu2023chiral} it has been shown that the unprojected fPEPS becomes chiral from $M\ge 2$, and the correlation functions improve quantitatively with increasing $M$. However, general topological properties of the  fPEPS remain unclear after Gutzwiller projection, and are not accessible (except for the ES) by conventional PEPS techniques. Hence we introduce the following Monte-Carlo method to probe the properties of Gutzwiller projected Gaussian fPEPS.

\subsection{Monte-Carlo technique}

The Gutzwiller projected fPEPS wavefunctions discussed previously are analysed within a standard Markov chain Monte Carlo framework \cite{becca2017}. In particular, overlaps between two projected wavefunctions $\ket{\psi}$ and $\ket{\phi}$ can be computed straightforwardly as follows,
\begin{equation}
    \frac{\braket{\psi|\phi}}{\braket{\psi|\psi}} =  \frac{ \sum_x \braket{\psi|x} \braket{x|\phi}}{\braket{\psi|\psi}} =   \frac{ \sum_x \lvert\braket{\psi|x}\rvert^2 \frac{\braket{x|\phi}}{\braket{x|\psi}}}{\braket{\psi|\psi}}
\end{equation}
where $\{\ket{x} \}$ is chosen to be the $S_z$ basis to enforce the one fermion per site constraint exactly. In this paper, we remain in the $S_z = 0$ sector, with equal number of up and down spins. Then, by sampling the normalized probability distribution
\begin{equation}
    P(x) = \frac{\lvert\braket{\psi|x}\rvert^2}{\braket{\psi|\psi}}
\end{equation}
one can estimate the wavefunction overlap as 
\begin{equation}
    \frac{\braket{\psi|\phi}}{\braket{\psi|\psi}} \sim \frac{1}{n}\sum_{i = 1}^n \frac{\braket{x_i|\phi}}{\braket{x_i|\psi}}
\end{equation}
where $n$ is the number of Monte Carlo runs, and $\{ \ket{x_i} \}$ are the spin configurations sampled in the Markov chain. We note that the cost of computing overlaps for the projected fPEPS is \textit{independent} of bond dimension, as the set of single-particle orbitals in real space can be obtained analytically for any $M$. This enables the calculations in section \ref{sec3vmc}, where we quantitatively analyze the fPEPS for $M=1\hdots6$.

\subsection{Variational iPEPS method}
In Section \ref{sec4ipeps}, we perform a variational study of the chiral Heisenberg antiferromagnetic model, taking the projected fPEPS parton ansatz as the initial state in our optimization. Firstly, we construct the Gutzwiller projected tensor for parton ansatz following Refs. \cite{li2023u,niu2023chiral}, where a single tensor of bond dimension $4^M$ contains two physical sites and satisfies $\rm U(1)\times \rm SU(2)$ symmetry. Secondly, we choose this tensor as initial state and variationally optimize the tensor elements with the $\rm U(1)\times \rm SU(2)$ virtual symmetry kept. To optimize the tensor, we adopt the automatic difference method \cite{liao2019differentiable} and choose the energy of the chiral spin model as the cost function. The energy is evaluated from the corner transfer matrix renormalization group (CTMRG) \cite{ctmrg1,ctmrg2} method, where the approximate contraction is controlled by the environment bond dimension $\chi$, and becomes exact in the
$\chi\rightarrow \infty$ limit.  

\section{Characterizations of projected fPEPS parton ansatz: VMC studies} 
\label{sec3vmc}

Several VMC algorithms have been developed to study topological properties of spin liquids, including entanglement entropy, modular matrices, and topological degeneracy \cite{zhang2011topological,yizhang2012,mei2015modular}. 
In this section, using VMC calculations on finite tori, we investigate the properties of the fPEPS wavefunctions, and compare them to the exact CSL state constructed from the parton ansatz \eqref{eq:corr_state}. We demonstrate that the fPEPS at finite bond dimension can capture the correct properties of the CSL. 
The gaussian fPEPS tensor is determined from optimizing the mean-field Hamitonian \eqref{eq:Hofstadter} on a $80\times 80$ torus. Subsequently, we put the optimized tensor on smaller $L\times L$ clusters to construct the many-body wave functions which are input to the Monte Carlo algorithm.

\subsection{Wavefunction fidelity}

\begin{figure}[H]
\includegraphics[width=0.9\columnwidth]{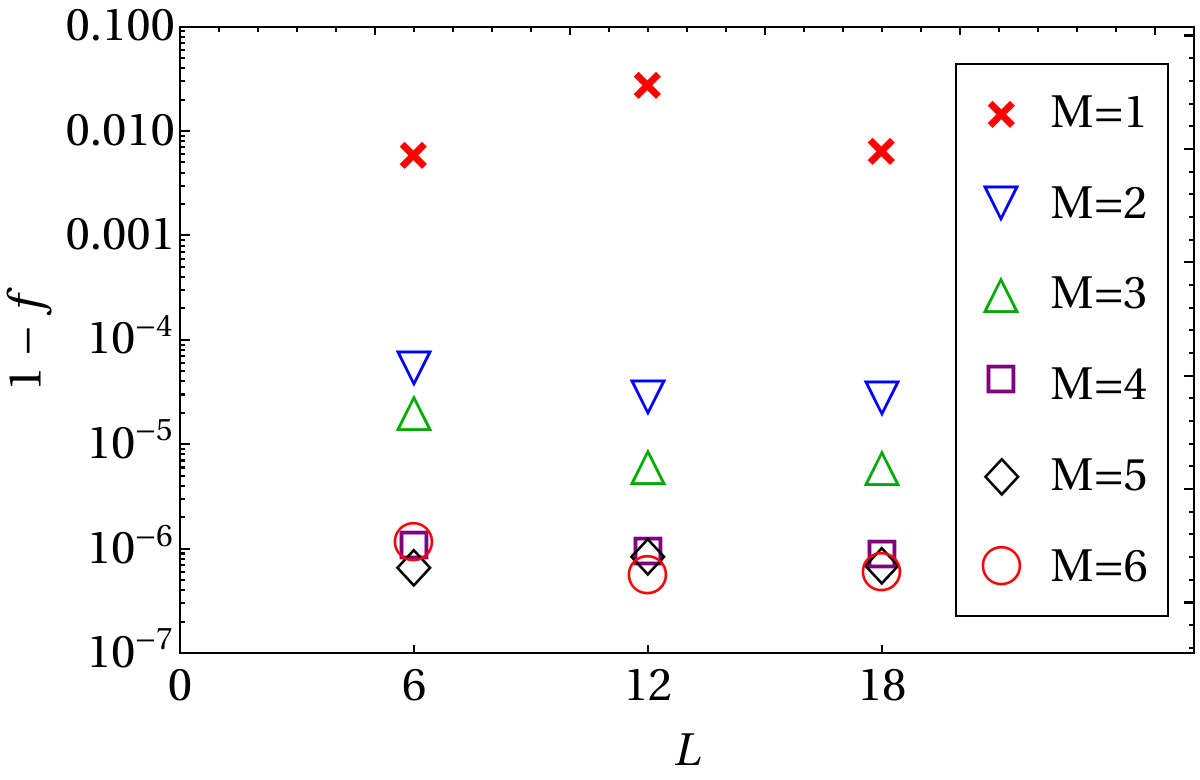}
\caption{\label{fig:fidelity} 
Infidelity $1-f$ plotted in logarithmic scale as a function of the system size $L$ for fPEPS states with $M = 1 \hdots 6$ and periodic boundary conditions.
The estimated error bars are smaller than the size of the symbols.}
\end{figure}

We first compute the normalized overlap between the projected exact CSL and the fPEPS states with periodic boundary conditions (PBC-PBC), given by
\begin{equation}
  O_M = \frac{\lvert \braket{\Psi_{\text{exact}}|\Psi_M} \rvert}{\sqrt{\braket{\Psi_{\text{exact}}|\Psi_{\text{exact}}} \braket{\Psi_{M}|\Psi_{M}}}}.
\end{equation}
By contracting physical indices of the PEPS, the overlap can be mapped to a partition function of a two-dimensional classical statistical model, thus decaying exponentially with system size.
We can then define the fidelity per unit area (free energy) $f = (O_M)^{1/L^2}$, which should show weak size dependence and converge to a finite value in the $L\rightarrow\infty$ limit.
The infidelity $1-f$ plotted in figure \ref{fig:fidelity} confirms these expectations. In addition, the diminishing infidelity with increasing $M$ clearly demonstrates the improving accuracy of the optimized fPEPS states. We note that similar results have been obtained for the other three choices of boundary conditions, i.e., PBC-APBC, APBC-PBC and APBC-APBC. 

\subsection{Spin-spin correlations}

\begin{figure}[H]
\hspace{-0.5cm}
\includegraphics[width=0.9\columnwidth]{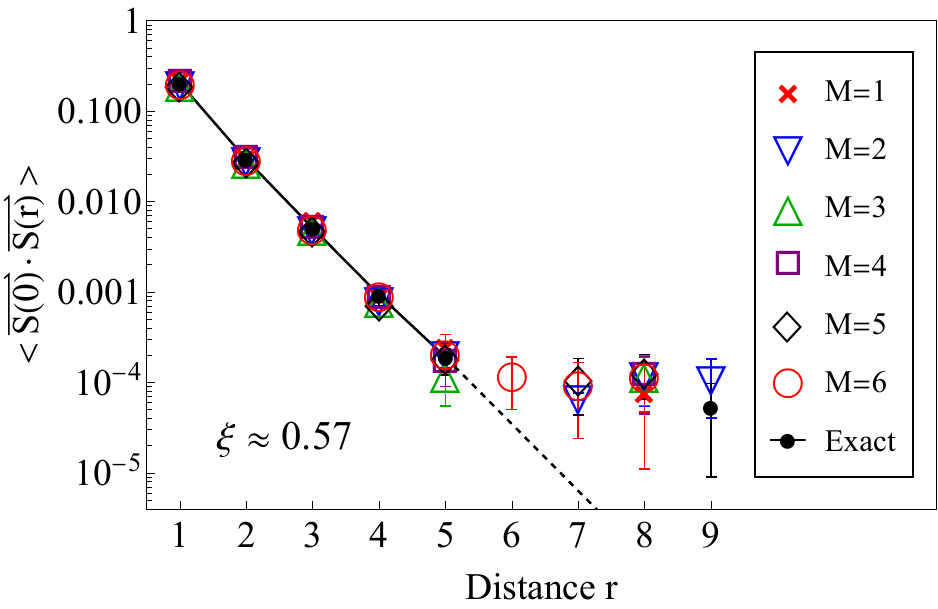}
\caption{\label{fig:spin_spin corr} 
Logarithmic plot of the magnitude of spin-spin correlations as a function of distance for the exact CSL and projected fPEPS, for the $L=18$ cluster with periodic boundary conditions. Symbols are ommited if the obtained value is lesser than one standard deviation of error. The expected correlations of the exact CSL at $r>5$ (in the thermodynamic limit) are shown as a dotted line, extending the exponential behavior.
}     
\end{figure}

To further confirm that the projected fPEPS describe the correct physical properties of the CSL, we compute the real space spin-spin correlations for the $L=18$ system, shown in figure \ref{fig:spin_spin corr}. We observe exponential behaviour at short distances as expected for a gapped state, with a very short correlation length $\xi \approx 0.57$. Further, the fPEPS states (for all values of $M$) are essentially indistinguishable from the exact CSL in terms of the correlations. 
Note that the saturation of the decay of the long-distance correlations for $r>5$ can be simply attributed to a finite size effect with periodic boundary conditions when $r\sim L/2$. Hence, we cannot definitively establish the presence of a 'gossamer tail', an artifact due to the bulk-boundary correspondence that has been discussed in several previous works \cite{chen2018non,hasik2022,niu2023chiral}.

\subsection{Topological properties: ground state degeneracy}

A fundamental characteristic of a topological ordered phase is a ground state degeneracy which depends on the topology of space \cite{wen90}. 
As mentioned previously, the exact $\rm SU(2)_1$ CSL state \eqref{eq:corr_state} has a two-fold degeneracy on a torus: imposing different boundary conditions on the parton ansatz before projection yields degenerate states that cannot be distinguished by local observables (like spin-spin correlations) after projection. In the thermodynamic limit the four states only span a two dimensional linear space \cite{yizhang2012}.

\begin{figure}[t]
  \centering
\subfigure{\includegraphics[scale=0.34]{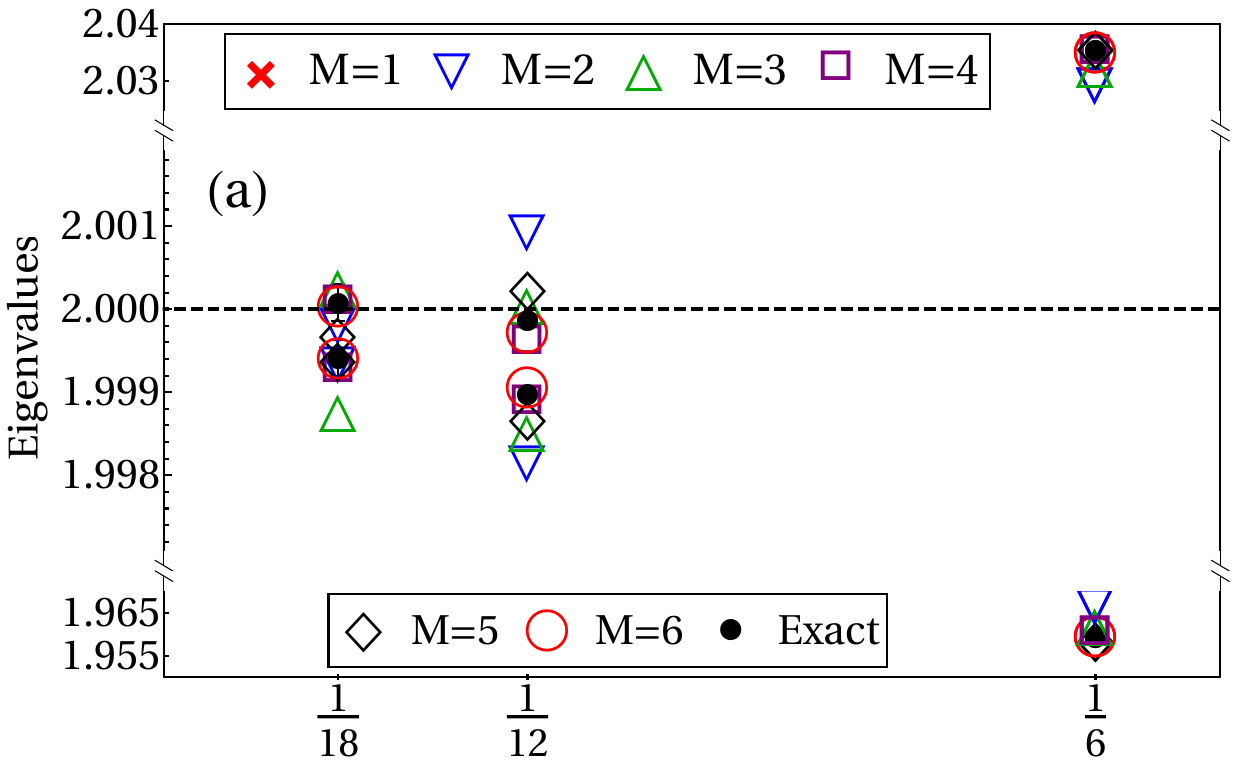}}\\
    \hspace{-0.4cm}
  \subfigure{\includegraphics[scale=0.34]{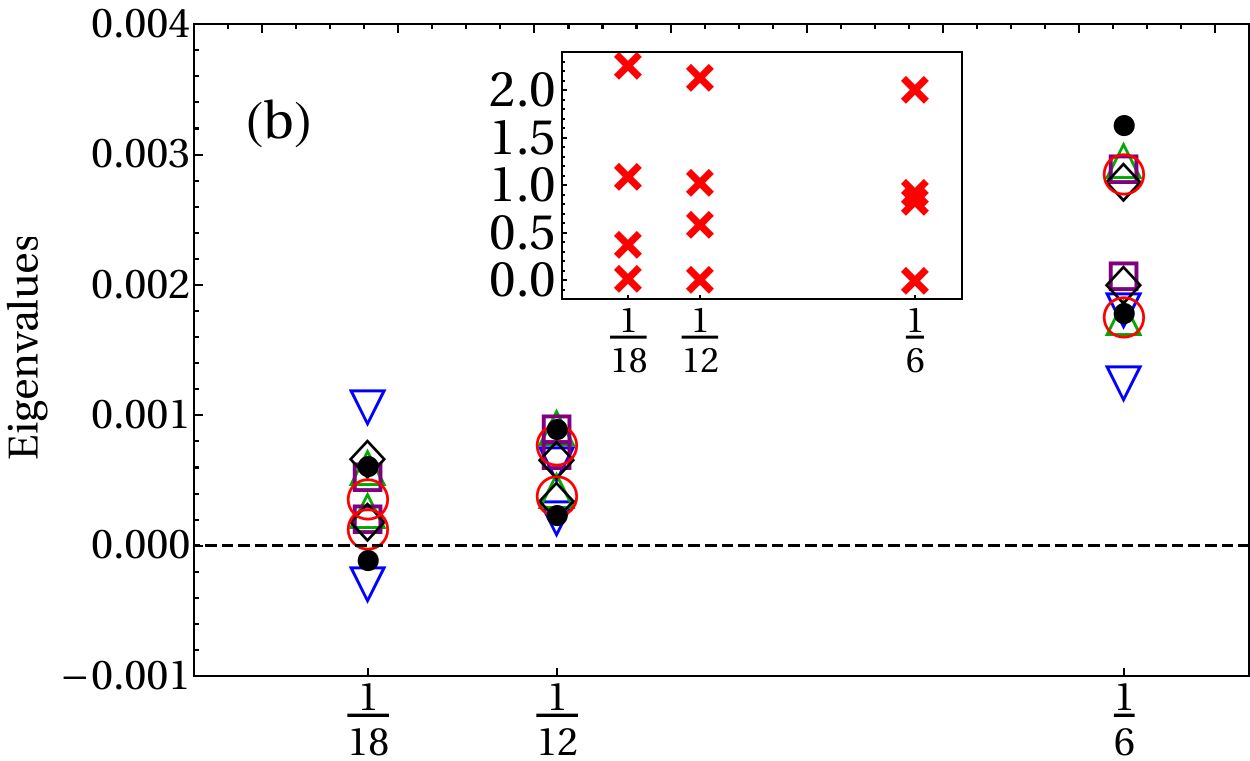}}
  \caption{Eigenvalues of the overlap matrix as a function of inverse system size $1/L$ for the exact and the fPEPS states. Pairs of eigenvalues converge to +2 and 0 in (a) and (b) respectively. The $M=1$ data is shown in the inset since it has a much larger deviation than the other fPEPS states. The estimated error bars are roughly the size of the symbols.}
  \label{fig:evals_vs_l}
\end{figure}

To investigate this property of the fPEPS and exact states on finite clusters, we compute the $4\times 4$ overlap matrix $O$ with elements 
\begin{equation}
O_{\alpha, \beta} = 
\frac{\braket{{\Psi}^{\beta}_M | {\Psi}^{\alpha}_M}}{\sqrt{\braket{{\Psi}^{\alpha}_M | {\Psi}^{\alpha}_M}\braket{{\Psi}^{\beta}_M | {\Psi}^{\beta}_M}}}
\end{equation}
where $\alpha$ and $\beta$ denote the four choices of boundary conditions on the torus. The rank of this hermitian matrix (which denotes the number of linearly independent eigenvectors) is the number of non-zero eigenvalues. In the thermodynamic limit, the eigenvalues of the exact state must converge to $\{+2,+2,0,0\}$ (the trace of the matrix being 4). In figure \ref{fig:evals_vs_l}, we plot the eigenvalues for both the projected fPEPS and the exact states as a function of $L$.  Remarkably, we observe that for \textit{fixed} $M > 1$, the eigenvalues converge to the exact result with increasing system size (already at $L=18$, the deviation is at most $10^{-3}$). 
In addition, from the analysis of the eigenvectors of the overlap matrix, we  have obtained the following relations (up to a gauge degree of freedom)
\begin{align}
 \begin{aligned}
    |\Psi^{\text{PBC-PBC}}\rangle &=\frac{|\Psi^{\text{PBC-APBC}}\rangle+|\Psi^{\text{APBC-PBC}}\rangle}{\sqrt{2}}, \\
    |\Psi^{\text{APBC-APBC}}\rangle &=\frac{|\Psi^{\text{PBC-APBC}}\rangle-|\Psi^{\text{APBC-PBC}}\rangle}{\sqrt{2}}
 \end{aligned}
 \end{align}
 with a very good accuracy whenever $L\ge 18$, consistent with the fact that the states on the RHS and the LHS of the above equations belong to the same IRREPs of the $C_4$ rotation group on clusters with $L_x=L_y$ (respectively even and odd). 
These results substantiate that fPEPS, even at finite bond dimensions, can accurately capture the correct topological degeneracy of CSL states.
This is one of the main results of our work.

\section{Simulation of the chiral $J_1-J_2-J_{\chi}$ model}
\label{sec4ipeps}
From the VMC analysis in the previous section, we have seen that projected (Gaussian) fPEPS can describe topological properties of CSL faithfully. However, for the purpose of studying frustrated spin models, the conventional parton ansatz has limited number of variational parameters such as hopping coefficients, Jastrow factors, etc.. On the contrary, PEPS can represent generic interacting states with the systematic increase of bond dimension. Now we conduct a variational PEPS study on the chiral $J_1-J_2-J_{\chi}$ model using projected GfPEPS as the initial ansatz. The spin-$1/2$ hamiltonian is given by
\begin{equation}\label{eq:hamiltonian}
{\cal H} = J_1 \sum_{\langle i,j \rangle} {\bf S}_i \cdot {\bf S}_j + J_2 \sum_{\langle \langle i,k \rangle \rangle} {\bf S}_i \cdot {\bf S}_k + J_{\chi} \sum_{\triangle_{ijk}} \left({\bf S}_i \times {\bf S}_j \right) \cdot {\bf S}_k.
\end{equation}
This model was initially proposed in \cite{Nielsen2013}, where exact diagonalization techniques revealed that the ground state had a very high overlap with the exact Kalmeyer-Laughlin state. The model exhibits a chiral spin liquid (CSL) phase over a large region of its phase diagram, and we select a specific set of parameters also discussed in that work. The Hamiltonian can be undersdood as an effective Hamiltonian of the Fermi-Hubbard model through the Schrieffer–Wolff transformation. Realizing the Fermi-Hubbard model in cold atom experiments remains an active area of research \cite{TARRUELL2018365, Wang2022,spar22}. Further, this model has been investigated in Ref. \cite{poilblanc2017}, where the sum of four triangular $J_{\chi}$ terms inside a plaquette has been equivalently written as a spin cyclic permutation $i(P_{ijkl}-P_{ijkl}^{-1})$ term. Ref. \cite{poilblanc2017} performed a variational study of this model with bosonic iPEPS, and found a regime of $\rm SU(2)_1$ CSL in the phase diagram. The optimized bosonic iPEPS provides good variational energy and correct level counting in the ES predicted by $\rm SU(2)_1$ WZW CFT. However, there exists a redundant chiral branch in the odd (semion) sector \cite{poilblanc2017,hackenbroich2018interplay} which contradicts both the theoretical prediction and recent numerical results obtained from DMRG on finite cylinders \cite{Yang2024,Zhang2024}. We emphasize that such artificial replication of chiral branches in bosonic iPEPS is quite general and is also found in the cases of $\rm SU(N)$ and non-Abelian CSLs \cite{chen2018non,chen2020,chen2021abelian,niu2022chiral}. 

On the other hand, in Ref. \cite{niu2023chiral} it was shown that the projected fPEPS from parton construction not only has correct level counting, but also has exact branch numbers in each topological sector of the ES. However, it is not clear whether such ES degeneracy of fPEPS is a robust or fine-tuned feature. From the variational study below we would like to show that (i) the parton state provides an energetically good initial guess for variational optimization and (ii) the optimized fPEPS state (which is beyond projected Gaussian states) still gives the correct ES counting and number of chiral modes, implying that our family of fPEPS states are not fine-tuned and provide a faithful description of CSLs.

\begin{figure}[H]
\includegraphics[width=\columnwidth]{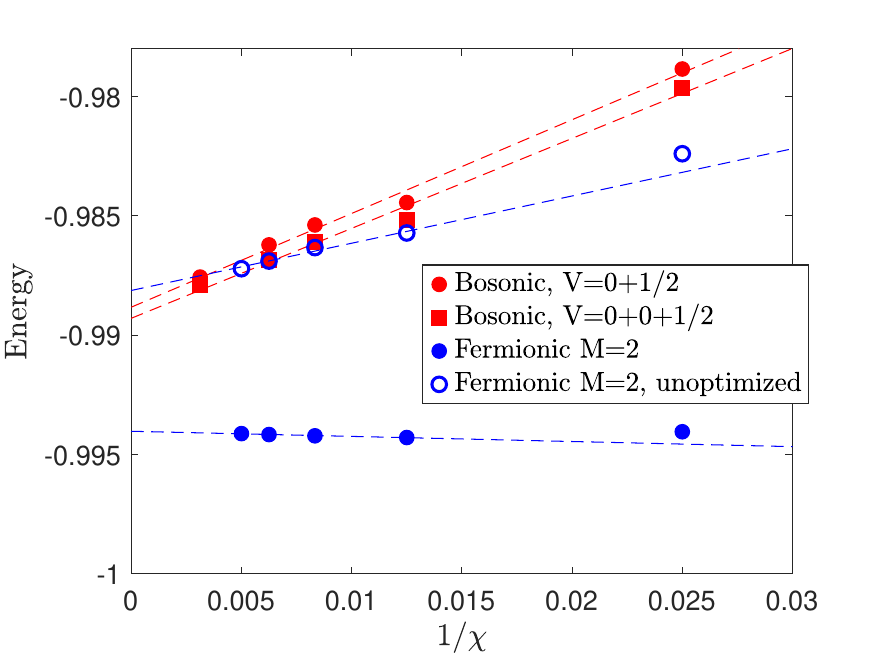}
\caption{\label{fig:var_opt} 
Variational energy of optimized bosonic iPEPS (red) and projected fermionic iPEPS (blue) for the spin model. Blue open circle shows the energy of the fermionic parton ansatz at $t_1=1,t_2=0.5$ without variational optimization of tensor elements.  
}
\label{fig:energy}
\end{figure}

\subsection{Variational energy}

We choose the parameters of the spin model as $ J_1=2\cos(0.06\pi)\cos(0.14\pi)$, $ J_2=2\cos(0.06\pi)\sin(0.14\pi)$, $J_{\chi}=4\sin(0.06\pi)$ which has been considered in Refs. \cite{Nielsen2013,poilblanc2017,hasik2022} and is known to be deep inside the CSL phase with the ground state energy being $E\approx  -1$.  We import the $\rm SU(2)$ bosonic iPEPS method there and compute variational energies as a reference. For the bosonic iPEPS, the tensor has $\rm SU(2)$ symmetry and the unit-cell size is chosen to be one such that each tensor only has one physical leg. The results for virtual spaces $V=0\oplus 1/2$ ($D=3$) and $V=0\oplus 0\oplus 1/2$ ($D=4$) are given in Fig. \ref{fig:energy}. As we extrapolate to the infinite $\chi$ limit the energies are around $-0.99$. 

In our fPEPS treatment, the initial parton state is still chosen at the hopping parameter $t_1=2t_2$ which corresponds to the largest band gap. Since the smallest bond dimension $M=1$ is non-chiral, we take $M=2$ ($D=16$) GfPEPS with Gutzwiller projection and compute the energy with respect to the spin model. Due to the gauge choice in the parton construction, the smallest unit-cell size is 2-sites along $x$ direction. As can be seen in Fig. \ref{fig:energy}, the unoptimized parton state already has good energy close to the optimized bosonic iPEPS at $D=3,4$. After optimizing the fPEPS tensor elements, the energy further improves to $E\approx -0.995$. When comparing the energies of bosonic and fermionic iPEPS, one should recall that since our fPEPS has two-sites unitcell in $y$ direction, the effective bond dimension in $y$ direction is $D_{\rm eff}=4$ per site, while the $x$ direction bond is much larger than that of bosonic iPEPS. This is consistent with the fact that our fermionic iPEPS has better energy.

\begin{figure}[H]
\includegraphics[width=\columnwidth]{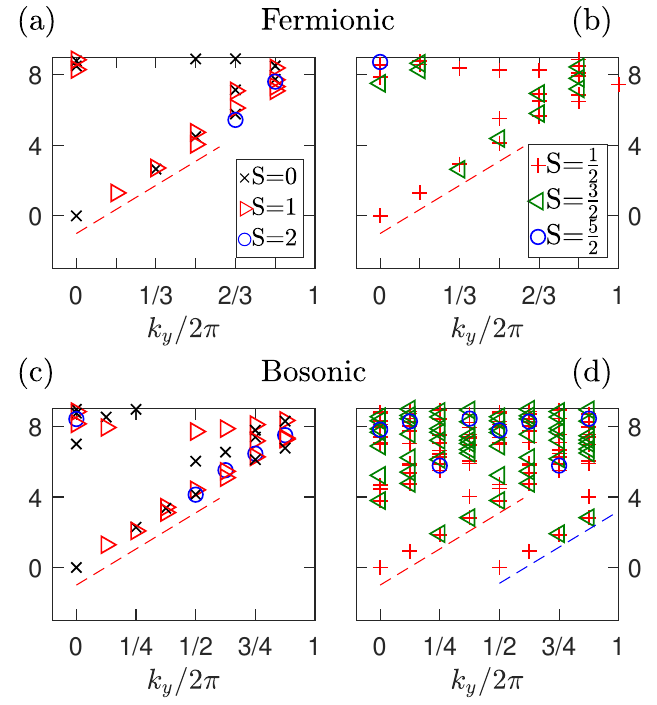}
\caption{ 
ES of optimized $M=2$ projected fermionic iPEPS (top) and $D=3$ bosonic iPEPS (bottom) for the spin model on cylinders of finite width 6 and 8, respectively.  Left and right columns correspond to even (integer spin) and odd (half-integer spin) sectors. The red dashed lines denote the theoretically predicted linear dispersions of the tower of states, while the blue dashed line denotes the redundant branch in the odd sector of bosonic iPEPS. 
}
\label{fig:ES}
\end{figure}

\subsection{Entanglement spectrum}
Aside from modular matrices, bulk topology of CSLs can also be characterized by edge (entanglement) spectrum according to bulk-boundary correspondence. To analyse the level counting of ES, we put our optimized tensor on a finite width cylinder and compute the ES of such translation-invariant state. To compute ES, the boundary Hamiltonian (transfer matrix fixed points) can be constructed exactly by exact contraction ~\cite{cirac2011entanglement} or approximately by grouping CTMRG environment $T$ tensors~\cite{poilblanc20162}. In the case of fermionic iPEPS, the tensor has $\rm U(1)\times \rm SU(2)$ virtual symmetry and the transfer matrix fixed points are labeled by virtual charge and parity of virtual spin, while in the case of bosonic iPEPS the tensor only has $\rm SU(2)$ virtual symmetry and the transfer matrix fixed points are labeled only by parity of virtual spin. Fig. \ref{fig:ES} (a)-(b) show ES of optimized $M=2$ fermionic iPEPS computed from CTMRG with $\chi=110$. The low energy spectrum shows $\rm SU(2)$ multiplets with counting $0,1,0+1,0+1+1,...,$ in the integer spin sector and $1/2,1/2,1/2+3/2,...,$ in the half-integer spin sector, satisfying prediction from $\rm SU(2)_1$ WZW CFT. Note that in both even and odd sectors there is only one low energy chiral branch which matches with recent DMRG results on finite cylinders~\cite{Yang2024,Zhang2024}. On the contrary, Fig. \ref{fig:ES} (c)-(d) show ES of $D=3$ bosonic iPEPS where the level counting is correct but an anomalous  
identical branch appears in the odd sector with a $\pi$-momentum shift. We also confirmed that the redundant branch can not be eliminated or shifted by increasing the bond dimension of the bosonic iPEPS. In future works, it would be interesting to see whether such artifact of bosonic iPEPS can be eliminated by further imposing virtual $\rm U(1)$ symmetry and/or increasing the unit-cell size like in the fPEPS case.

\section{Conclusion and outlook} 
Tensor Network methods have been widely applied to study frustrated spin models hosting possible spin liquid phases. However, due to a no-go theorem \cite{dubail2015tensor}, doubts have arisen regarding their ability to describe chiral spin liquid states. In this work, we investigated the topological properties of projected fPEPS ansatz for a CSL state. By performing a VMC analysis on the initial parton state, we demonstrated that fPEPS at finite bond dimension can accurately capture the topological GS degeneracy of CSLs. This achievement has been challenging with conventional bosonic PEPS due to the high computational cost involved \cite{poilblanc2015chiral}. Additionally, our results show that the fPEPS ansatz effectively captures the spin-spin correlation functions of the CSL. Given the high quality of overlap fidelity, we anticipate that fPEPS will also accurately reproduce the modular S and T matrices.

Further, non-Gaussian fPEPS ansatze were optimised variationally for the $J_1 - J_2 - J_\chi$ Heisenberg model and were found to have competitive energy compared to their bosonic counterparts, while retaining the correct level counting of the entanglement spectrum as opposed to bosonic iPEPS which have a duplicate branch in the odd topological (semion) sector. 

Our work provides further supporting evidence for using PEPS methods in describing topologically ordered states.
For example, fPEPS ansatz can further be used to study fermionic Hofstadter-Hubbard model~\cite{Tu2018} where both the Chern insulating phase and CSLs can be tuned by varying the strength of Hubbard interaction (e.g. Ref.~\cite{kuhlenkamp2022tunable}). In that case, our Gaussian fPEPS ansatze with partially projected doublons are expected to be good initial variational ansatze at finite Hubbard interaction.

\section*{Acknowledgements}

We thank F. Becca and Y. Iqbal, for discussions on VMC, as well as J.-W. Li and J.-Y. Chen. S.B acknowledges financial support from the Indo-French Centre for the Promotion of Advanced Research - CEFIPRA Project No. 64T3-1. D.P. acknowledges support from the TNTOP ANR-18-CE30-0026-01 grant awarded by the French Research Council. This work was granted access to the HPC resources of CALMIP center under allocations 2023-P1231 and 2024-P1231. 


%

\end{document}